\documentclass[useAMS,usenatbib]{mn2e}

\newcommand{\swift}{{\it Swift}}
\newcommand{\xmm}{{\it XMM-Newton}}

\newcommand{\kepler}{{\it Kepler}}

\newcommand{\rxte}{{\it RXTE}}
\newcommand{\ginga}{{\it Ginga}}

\usepackage{graphicx}
\usepackage{epstopdf}

\title[PDSs of \kepler\ AGN observations]{Multicomponent Power Density Spectra of \kepler\ AGNs, an instrumental artifact or a physical origin?}
\author[A. Dobrotka et al.]{
A. Dobrotka$^1$\thanks{E-mail: andrej.dobrotka@stuba.sk}, P. Bez\'ak$^{1}$, M. Revalski$^{2,\dagger}$ and M. Str\'emy$^{1}$\\
$^1$Advanced Technologies Research Institute, Faculty of Materials Science and Technology in Trnava, Slovak University of Technology\\
in Bratislava, J. Bottu 25, 917 24 Trnava, Slovakia\\
$^2$ Department of Physics and Astronomy, Georgia State University, 25 Park Place, Suite 605, Atlanta, GA 30303, USA\\
$^\dagger$ National Science Foundation Graduate Research Fellow\\
}

\begin{document}

\date{\today}

\pagerange{\pageref{firstpage}--\pageref{lastpage}} \pubyear{2018}

\maketitle
\label{firstpage}

\begin{abstract}
We analysed the light curves of four active galactic nuclei (AGN) from the \kepler\ field, and find multicomponent power density spectra with characteristic frequencies that are surprisingly similar to other \kepler\ AGNs (including ZW229-15). An identical time series analysis of randomly selected planet candidate stars revealed the same features, suggesting an instrumental origin for the variability. This result is enigmatic, as these signals have been confirmed for ZW229-15 using independent observations from \swift. Based on our re-analysis of these \swift\ data and test simulations, we now distinguish the instrumental artifact in \kepler\ data from the real pattern in \swift\ observations. It appears that some other AGNs observed with instruments such as \xmm\ show similar frequency components. This supports the conclusion that the similarity between the variability timescales of the \kepler\ artifact and real \swift\ features is coincidental.
\end{abstract}


\section{Introduction}

\label{introduction}

Time-domain studies of Active Galactic Nuclei (AGN) yield valuable information on the physical mechanisms responsible for the variability of these compact objects. The improved temporal coverage provided by the \kepler\ mission (\citealt{borucki2010}, \citealt{koch2010}) afforded a unique opportunity for nearly uninterrupted multi-year observations. Significant work has been done in this area, with the focus on fourier and Power Density Spectra (PDS) analyses (\citealt{mushotzky2011}, \citealt{carini2012}, \citealt{wehrle2013}, \citealt{edelson2014}, \citealt{revalski2014}, \citealt{chen2015}, \citealt{kasliwal2015a}, \citealt{kasliwal2015b}, \citealt{shaya2015}, \citealt{mohan2016}, \citealt{dobrotka2017}).

It is widely accepted that AGN are powered by material from an accretion disk that falls onto a central supermassive black hole, creating a source of fast stochastic variability, called flickering. This variability manifests itself as red noise in PDS that are calculated from photometric data, creating a decreasing linear trend in a log-log periodogram. However, some additional patterns such as break frequencies or Lorentzian components suggest that characteristic frequencies are finger-prints of underlying physical processes. The search for these PDS components allows for a better understanding of the accretion physics, and data with adequate temporal characteristics are needed for such studies. The best instrument combining quality and quantity of data is the \kepler\ satellite, with almost continuous observations over a duration of approximately 1200 days.

The first study of AGNs observed by \kepler\ was performed by \citet{mushotzky2011}. These authors studied the PDS slopes of four objects and did not detect any characteristic break frequency. Subsequent re-analysis of one of the four objects ZW229-15 performed by \citet{edelson2014} revealed a break frequency at $\sim 5$\,days. PDSs of another four AGNs in \kepler\ field studied by \citet{wehrle2013} and \citet{revalski2014} also did not reveal long term break frequencies.

\citet{dobrotka2017} re-analysed the light curves of four AGNs studied by \citet{mushotzky2011} and detected multicomponent PDSs for all four AGNs with two very similar break frequencies approximately at log($f$/Hz) = -5.2 and -4.7 in all four cases. The authors used also \swift\ data of ZW229-15 and confirmed both frequency components in X-rays. This is an important result, because it not only localises the source, but also confirms the reality of the signals. The latter is very important, because \kepler\ has many instrumental effects. For example, some spurious excess power in the power density spectra can be present (\citealt{edelson2014}), or the light curves of AGNs could/should be reprocessed (\citealt{kasliwal2015b}) to exclude non-AGN variability of the host galaxy. Other effects like moir\'e pattern can also significantly affect the photometry (\citet{kolodziejczak2010}).

However, \citet{smith2018} performed an extensive and detailed study of \kepler\ AGN, revealing many instrumental effects. Unfortunately, it appears that the signal at log($f$/Hz) = -5.2 is of instrumental origin, and it is present because of the \kepler\ thermal recovery period (see Fig.~7 of \citealt{smith2018}). This finding creates an enigma; is the signal real based on the \swift\ detection, or it is an artifact caused by the thermal recovery period?

The best way to test whether the detected variability is real and has a physical origin is to perform the same PDS analysis on different objects in nature and search for commons signals. In this paper we search for the same PDS components as found by \citet{dobrotka2017}, but using the same data as in \citet{revalski2014} (Section~\ref{pds}). Besides the \kepler\ data (Section~\ref{pds_kepler}) we also re-analyse the \swift\ data of ZW229-15 (Section~\ref{section_pds_swift}) already presented in \citet{dobrotka2017}. The Section~\ref{pds_kic} presents the PDS analysis of planet candidates that should have a completely different variable nature, and the results are discussed and compared to other PDS detections from the literature in Section~\ref{discussion}.

\section{Data}

In Table~\ref{objects} we summarize the properties of the four radio-loud AGN investigated by \citet{wehrle2013} and \cite{revalski2014}. We re-examine these data with the goal of further differentiating between instrumental artifacts and signals with true astrophysical origins. To accomplish this, we employed three different categories of data. First, we used the power density spectra (PDS) as calculated and presented in \citet{revalski2014}. The light curves for these PDSs were corrected to fill missing null points and monthly downlink gaps, subtract thermally induced focus shifts, and remove rare outliers. The flux levels between individual quarters were then multiplicatively matched, with and without end-matching applied to individual quarters. Additional details of target selection, data characteristics, and light curve processing are given in \citet{wehrle2013} and \cite{revalski2014}.
\setlength{\tabcolsep}{1.2pt}
\begin{table}
\caption{The four radio-loud \kepler\ AGN studied by \citet{wehrle2013} and \citet{revalski2014}.}
\begin{center}
\begin{tabular}{ccccc}
\hline
\hline
Object & Catalog & Kepler & Kepler & Redshift\\
Designation & Name & ID & Magnitude &\\
\hline
A & MG4\,J192325+4754 & 10663134 & 18.6 & 1.520\\
B & MG4\,J190945+4833 & 11021406 & 18.0 & 0.513\\
C & CGRaBS\,J1918+4937 & 11606854 & 17.8 & 0.926\\
D & [HB89]\,1924+507 & 12208602 & 18.4 & 1.098\\
\hline
\end{tabular}
\end{center}
\label{objects}
\end{table}

To ensure we retain all potential instrumental effects, and to have an equivalent comparison to the light curves in \citet{dobrotka2017}, we also downloaded the light curves for the AGN from \citet{wehrle2013} and \cite{revalski2014} and applied only minimum corrections (see below).

The last data category is comprised of randomly selected stars with planet candidates as the data for these objects should contain similar instrumental effects while having distinctly different origins of intrinsic variability. The primary motivation is to search for instrumental effects that may be convolved with true astrophysical variations.

\section{PDS Analysis}
\label{pds}

\subsection{Methodology}
\label{method}

The light curves downloaded from the \kepler\ archive were pre-processed using the \verb|hampel| utility within MATLAB to fill rare null points and replace outliers $>$ 3-$\sigma$ with local median. The thermal decline sections were excluded and adjacent light curves separated by gaps were end-matched so that the end points of each sub-sample transition smoothly to the starting points of the next sub-sample.

Any end-matching procedure supposes the same flux at the boundaries of the two adjacent light curve sub-samples, which is not correct. However, in \citet{dobrotka2017} we showed by simulation that the end-matching process does not significantly influence the resulting PDS. Therefore, any small imprecision is negligible, what is important in the case of long gaps.

Our new PDSs are calculated using the Lomb-Scargle algorithm (\citealt{scargle1982}) that can handle the gaps between individual quarterly light curves and non-equidistant data points. We first divided the light curves into 5 equal length sections, which we found to be the best balance between frequency resolution and noise in the PDSs. We then calculated the log-log periodogram for each section individually and averaged the results. We binned the periodograms in 0.05 logarithmic frequency steps\footnote{A minimum number of points (5) must be present in the bin, otherwise the bin is larger until the minimal number is reached.}, which is the same step size used in \citet{dobrotka2017}. As an indicator of the intrinsic scatter within each averaged frequency interval we calculated the standard error of the mean for the points in each bin.

The selection of the frequency interval and step depends on the characteristics of the sub-sample. The low frequency end and the frequency step of the PDSs are usually chosen according to the temporal extent of the light curve. The high frequency end is limited by white noise and is empirically limited to log($f/{\rm Hz}) = -4.0$. The ordinate axis is typically the power ($p$) or frequency multiplied by power ($f \times p$). The former is typically employed for slope analysis while the later visually highlights subtle PDS features. For the fitting procedures we used the {\small GNUPLOT}\footnote{http://www.gnuplot.info/} software. The fitting model consists of a linear and $n$ Lorentzian functions
\begin{equation}
P(f) = a + b F + \sum_{i=1}^{n} \left[ \frac{c_{\rm i} \Delta_{\rm i}}{\pi}\frac{1}{\Delta_{\rm i}^2 + (F - F_{\rm bi})^2} \right],
\label{model}
\end{equation}
where the fitting procedure we performed in logarithmic space, i.e. $P = {\rm log}(p)$ (or alternatively ${\rm log}(f \times p$)), $a$, $b$, and $c$ are constants, $F = {\rm log}(f)$ , $F_{\rm b} = {\rm log}(f_{\rm b})$ is the searched break frequency or PDS component and $\Delta$ is the half-width at half-maximum. The Lorentzian profile is useful to describe the individual "humps" in the PDS and the linear function describes the rising trend towards lower frequencies.

We focused our study on the same frequency interval as \citet{dobrotka2017}: from ${\rm log}(f/{\rm Hz}) = -6.0$ to the end of the second "hump". The latter is variable from case to case and not uniform because the model does not fit the white noise portion of the PDS. The low resolution PDSs were fit over larger frequency intervals.

\subsection{\kepler\ Results}
\label{pds_kepler}

\begin{figure}
\resizebox{\hsize}{!}{\includegraphics[angle=-90]{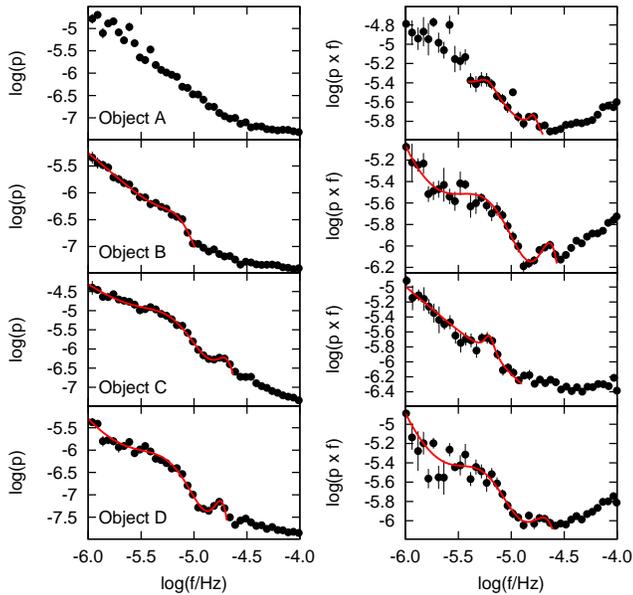}}
\caption{The PDSs (black points) with fits (red lines). In the left column are the original re-bined PDSs from \citet{revalski2014}. The right column shows our recalculated PDSs.}
\label{pds_all}
\end{figure}

We first re-analyzed the original PDSs from \citet{revalski2014} as presented in the left column of their Fig. 4. The results are shown in the left column of Fig.~\ref{pds_all}. The PDSs were re-binned as described above and the same multi-component shape reported in \citet{dobrotka2017} is visible, except for objects A and B. The former do not show any structure, while the latter show only the lower break frequency. Individual fits are presented in Fig.~\ref{pds_all} as red lines and the best fit parameters are listed in Table~\ref{pds_param}.

The PDSs of the re-constructed light curves with fits are shown in the right panel of Fig.~\ref{pds_all}. Only the high frequency component in object C is missing, otherwise all PDSs show the same PDS components detected by \citet{dobrotka2017}. The fitting of object A was problematic due to the sporadic points at log($f$/Hz) = -5.0. As the trend is visually clear, we excluded this point from the fitting procedure.
\begin{table}
\caption{List of PDS parameters as lower ($f_{\rm b1}$), higher ($f_{\rm b2}$) with standard errors. Reduced $\chi^2$ and degrees of freedom (dof) are also listed. $n$ is the number of Lorentzian components. For each object the first row lists the fit results for the ``original'' re-analyzed PDSs from \citet{revalski2014}, while the second shows our "new" recalculations.}
\begin{center}
\begin{tabular}{llccccr}
\hline
\hline
object & & n & log($f_{\rm b1}/{\rm Hz}$) & log($f_{\rm b2}/{\rm Hz}$) & $\chi^2_{\rm red}$ & dof\\
\hline
A & original & -- & -- & -- & --\\
 & new & 2 & $-5.22 \pm 0.03$ & $-4.79 \pm 0.01$ & 0.32 & 5 \\
B & original & 1 & $-5.15 \pm 0.07$ & -- & 0.69 & 15\\
 & new & 2 & $-5.02 \pm 0.24$ & $-4.53 \pm 0.07$ & 0.99 & 21\\
C & original & 2 & $-5.15 \pm 0.24$ & $-4.66 \pm 0.09$ & 0.86 & 19\\
 & new & 1 & $-5.20 \pm 0.01$ & -- & 0.59 & 17\\
D & original & 2 & $-5.08 \pm 0.42$ & $-4.70 \pm 0.07$ & 2.16 & 18\\
 & new & 2 & $-5.18 \pm 0.15$ & $-4.55 \pm 0.10$ & 1.71 & 20\\
\hline
ZW229-15 & & 2 & $-5.24 \pm 0.08^2$ & $-4.65 \pm 0.11^2$ & $0.99^2$ & $15^2$\\
& \swift\ & -- & $-5.48 \pm 0.20^2$ & $-4.65 \pm 0.15^2$ & $3.79^2$ & $20^2$\\
& \swift\ & 2 & $-5.57 \pm 0.14^3$ & $-4.56 \pm 0.11^3$ & $0.69^3$ & $8^3$\\
& \swift\ & 3 & -- & $-4.46 \pm 0.14^4$ & $0.84^4$ & $35^4$\\
\hline
KIC10525077 & & 2 & $-5.10 \pm 0.03$ & $-4.70 \pm 0.02$ & 3.09 & 20\\
 & det. & 2 & $-5.12 \pm 0.01$ & $-4.702 \pm 0.004$ & 0.37 & 8\\
KIC3558849 & & 1 & $-5.15 \pm 0.03$ & -- & 2.92 & 12\\
 & det. & 2 & $-5.09 \pm 0.01$ & $-4.74 \pm 0.03$ & 1.61 & 4\\
KIC5437945 & & -- & -- & -- & -- & --\\
 & det. & 2 & $-5.14 \pm 0.01$ & $-4.70 \pm 0.03$ & 1.14 & 5\\
KIC5446285 & & 2 & $-5.19 \pm 0.05$ & $-4.61 \pm 0.06$ & 2.09 & 20\\
 & det. & 2 & $-5.21 \pm 1.45$ & $-4.65 \pm 0.15$ & 2.62 & 7\\
KIC8751933 & & -- & -- & -- & -- & --\\
 & det. & -- & -- & -- & -- & --\\
\hline
\end{tabular}
\end{center}
$^1$ From \citet{edelson2014}. $^2$ From \citet{dobrotka2017}.\\
$^3$ This work using constant frequency (in log) bins. $^4$ This work using constant number of points per PDS bin.
\label{pds_param}
\end{table}

\subsection{\swift\ Reanalysis of ZW229-15}
\label{section_pds_swift}

The presence of very similar PDS components in all eight \kepler\ AGNs from \citet{revalski2014} and \citet{dobrotka2017} implies the signals are instrumental in origin. Additional support for an astrophysical origin in \kepler\ data of ZW229-15 was presented in \citet{dobrotka2017} through additional PDS calculations using \swift\ data. We found the same PDS components, which supports the features having a physical origin. However, the similarity of these components to all eight aforementioned objects is suspicious and motivated us to re-analyze the ZW22-15 \swift\ data. In \citet{dobrotka2017} we did not divide the \swift\ light curve into sub-samples because of the quality and sparse coverage of the light curve. We did so in this paper by dividing into 2 sub-samples, and performed two different binning procedures, one based on constant frequency step and the second using a constant number of points per bin. The former is the same as presented in Section~\ref{method}, and in the latter we averaged 10 successive points in the log-log averaged periodogram. For the low frequency portion of the PDS we did not use the entire light curve due to the poor light curve coverage. We choose an interval with the best coverage between 711.424 and 764.815\,days (see Fig.~2 of \citealt{dobrotka2017}) instead, with a corresponding frequency of log($f$/Hz) = -6.66. This is the longest variability measurable and making real sense based on the temporal length of the most continual data segment.

\begin{figure}
\resizebox{\hsize}{!}{\includegraphics[angle=-90]{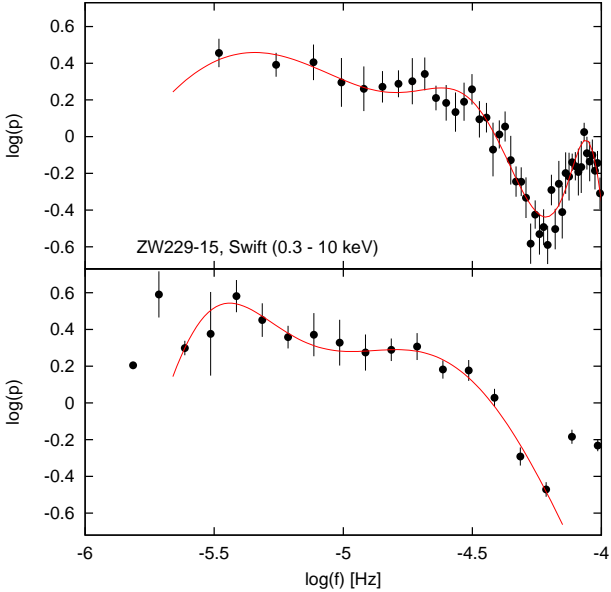}}
\caption{The recalculated PDS of ZW229-15 (black points with standard errors) from \citet{dobrotka2017} after division into two sub-samples using two binning procedures (see Section~\ref{method}). The best fit profile is shown in red.}
\label{pds_swift}
\end{figure}

The resulting PDSs are shown in Fig.~\ref{pds_swift} and the corresponding PDS component frequencies are given in Table~\ref{pds_param}. From the three component fit in the upper panel of Fig.~\ref{pds_swift} we read only the $f_{\rm b2}/{\rm Hz}$ value. The $f_{\rm b1}/{\rm Hz}$ consists of only a few points without any obvious PDS hump because of the low frequency resolution. Therefore, we do not take this value into account. One additional higher PDS component was added in order to get a smooth PDS description. This value near log($f$/Hz) = -4.1 is a result of variable sampling in the most covered interval between 711.424 and 764.815\,days (see Fig.~2 of \citealt{dobrotka2017}). The lower panel in Fig.~\ref{pds_swift} shows two well resolved and fitted hump features.

The high frequency PDS component detected in the \swift\ data differs slightly from the value detected in \citet{dobrotka2017}, but still agrees with the \kepler\ equivalent within the errors. However, the situation is considerably different in the low frequency case, where the new \swift\ fit yields a significantly lower value than in the \kepler\ case.

\begin{figure}
\resizebox{\hsize}{!}{\includegraphics[angle=-90]{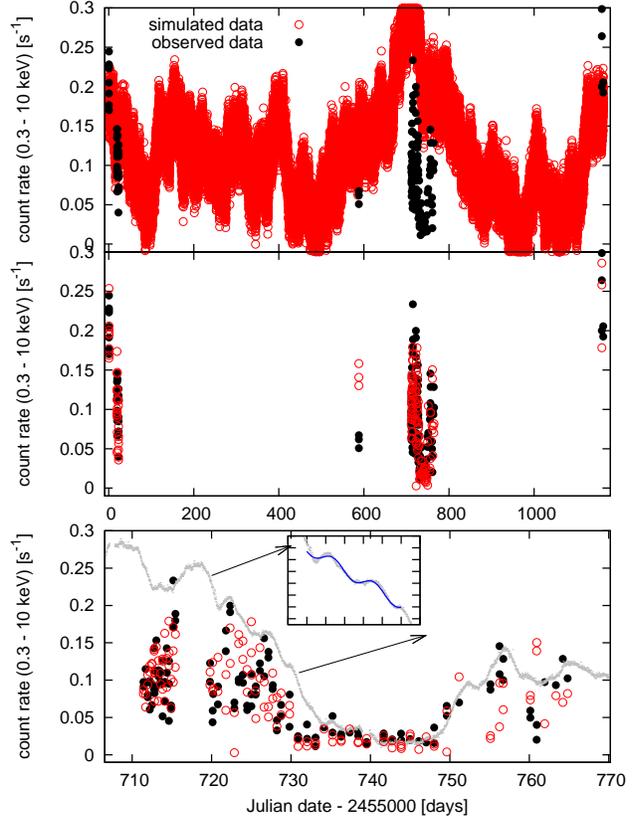}}
\caption{Comparison of observed \swift\ light curve of ZW229-15 with randomly selected example (top panel) and best example (middle and bottom panels) of simulated data (see text for details). The gray points are the vertically adjusted \kepler\ light curve points (figure reproduced from \citealt{dobrotka2017}). The bottom inset panel represents the sinusoidal fit with linear decreasing trend to the \kepler\ light curve subsample.}
\label{simul_lc_swift}
\end{figure}

To investigate the significance of both PDS components calculated from the \swift\ data we simulated noisy light curves using method by \citet{timmer1995}. This algorithm generates light curves with variabilities following an input PDS. For the latter, we first used power law fit describing simple red noise without the PDS components under investigation. The majority of observed points are localised between days 700 and 800 in Fig.~\ref{simul_lc_swift}, while the observations at the beginning and the end of the entire light curve have slightly higher flux. The latter have only a very minor influence on the result, but in order to perform the simulations rigorously we added this long term trend to the simulation, i.e. we added a parabolic trend taken from the fit to the observed data, and subsequently subtracted the mean of the fitted values, so as not to influence the observed mean value. We then simulated uninterrupted light curves with 1000\,s sampling with the first and last values taken from the \swift\ light curve. The mean flux and amplitude of variability\footnote{The so called $\sigma_{\rm rms}$ defined as square root of the variance.} was set equal to the observed data. Moreover, the bottom panel of Fig.~\ref{simul_lc_swift} shows the well-sampled portion of the observed light curve. A clear minimum is present. We removed these low flux data before calculating the observed mean flux and $\sigma_{\rm rms}$. Subsequently, we adapted the sampling of the entire simulation to match that of the \swift\ data by linear interpolation between each set of points, i.e. a new flux value with time coordinate of the \swift\ light curve is choosen using linear function between two surrounding simulated points. Any simulated negative count rates were returned to zero, and we visually adapted the clear brightness minimum in the data by multiplying the flux by a factor of 0.2.

\begin{figure}
\resizebox{\hsize}{!}{\includegraphics[angle=-90]{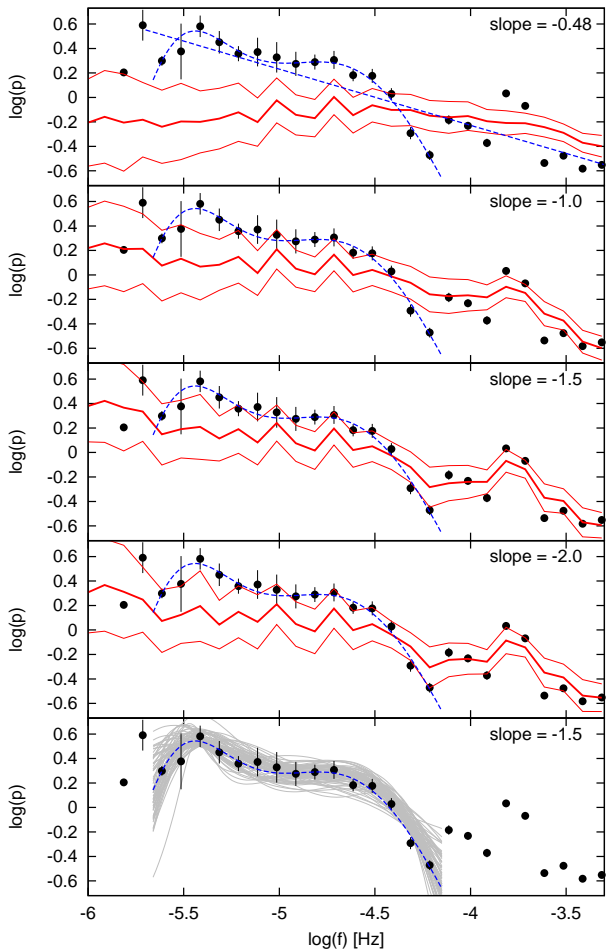}}
\caption{Mean values (thick red solid line) and 1-$\sigma$ interval (thin red solid lines) after one thousand simulations compared to the observed PDS (black points from lower panel of Fig.~\ref{pds_swift}). The blue dashed curve is the multicomponent fit, while the blue dashed line in the top panel represents the power law fit to the PDS describing a hypothetical red noise. The corresponding power law used for each simulation is labeled. The bottom panel compares the observed PDS and fit with selected fits to the simulations satisfying the criteria defined in the text.}
\label{simul_pds_swift}
\end{figure}

We calculated the corresponding PDSs in the same way as for the observed data. The mean value with a 1-$\sigma$ interval after one thousand simulations\footnote{We choose the constant frequency step for binning, because both PDS components are seen, and we expanded the frequency interval toward Nyquist frequency to see all patterns resulting from data sampling.} compared to the observed PDS is shown in top panel of Fig.~\ref{simul_pds_swift} with the power law fit (slope of -0.46). Clearly, the observed sampling strongly suppresses the red noise character\footnote{Keeping the light curve fully and equidistantly sampled, the simulations follow the fitted power law.}, with the observed PDS components significantly above the simulated power, which does not describe our results.

If the resampling reduces the red noise character, the original signal should have a steeper slope. We used three different values of -1.0, -1.5 and -2.0. For the reddest cases, another problem with long-term trend arises. The top panel of Fig.~\ref{simul_lc_swift} shows one simulation example with the reddest power law. The long-term trend clearly does not agree with the observed data. The poor coverage of \swift\ data does not allow us to distinguish whether the observed data are a random samples with low flux from a light curve with an underlying long-term trend, or whether the long-term trend is totally absent. Therefore, we extracted the long-term trend by fitting a 10$^{\rm th}$ order polynomial and added the parabolic trend. The best example of such a light curve\footnote{Selected as the one with minimal sum of squared residuals between observation and simulation.} from 1000 simulations using a power-law slope of -2 is shown in the middle and bottom panels of Fig.~\ref{simul_lc_swift}.

All simulations are shown in Fig.~\ref{simul_pds_swift}. Apparently, the highest slopes of -1.5 and -2.0 best describe the observed PDS, including the pattern around log($f$/Hz) = -3.8. The discussed $f_{\rm b2}$ signal (represented by the two component fit) is localised at the 1-$\sigma$ boundary, yielding a high probability that this signal is a result of a simple red noise and is not astrophysical. However, the $f_{\rm b1}$ signal is still above this, which implies its reality.

To quantify the probability that the low frequency component $f_{\rm b1}$ is instrumental, we fitted the simulated PDSs with the same multicomponent model (Eq.~\ref{model}) as the observed data. We selected fits with apparent two component shape and reaching power equal to the lower error bar of the highest PDS point at log($f$/Hz) = -5.41. The numbers are 24, 64 and 61 for red noise slopes of -1.0, -1.5 and -2.0, respectively. This yields probabilitites that the low frequency component $f_{\rm b2}$ is not real $\sim$ 2 and 6\%. The latter is valid for the two reddest cases. Taking the maximum power of the observed fit as a threshold criterium, instead of the lower error of the highest point, the probability is even lower.

\subsection{Comparison With Stellar Sources}
\label{pds_kic}

A clear way to test whether the detected variability is coming from the AGN or an instrumental artifact is to perform the same PDS analysis on completely different objects in nature. This is simple using \kepler\ data as the archive has numerous stellar sources. We randomly selected five sources with confirmed planets and performed the same PDS calculation.

The resulting PDSs are shown in the left column of Fig.~\ref{pds_kic}. Several PDSs clearly shows the same PDS components as the studied AGN. Not every case is clear and some objects show no common components. This could be due to large amplitude, long-term variability dominating the PDSs and damping faint structures. To correct for this we de-trended the light curves with a running median using 101 points, \footnote{The median is calculated from the point itself and 50 points on either side, yielding a median interval calculation of approximately 181800\,s. The corresponding frequency is log($f$/Hz) = -5.25, which is smaller than the potential signals such that any influence should be minimal.} and the resulting PDSs are shown in the right panel of Fig.~\ref{pds_kic}. The fits reveal obvious structures with peak frequencies summarized in Table~\ref{pds_param}. Only the last object does not show the humps, having only one or two points with any significant power excess that are not well fit with a Lorentzian profile.
\begin{figure}
\resizebox{\hsize}{!}{\includegraphics[angle=-90]{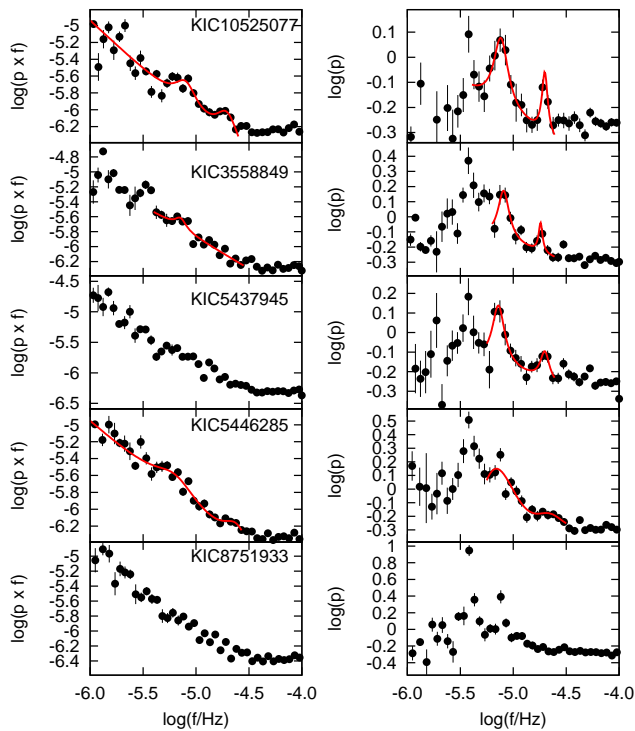}}
\caption{PDSs (points with vertical lines) with fits (red lines) of selected stars from \kepler\ field. Left column - high resolution PDSs, right panel - the same as left column but after de-trending using running median.}
\label{pds_kic}
\end{figure}

\section{Discussion}
\label{discussion}

We have in total eight AGN observations (\citealt{dobrotka2017}, and this work) by \kepler\ showing the same PDS components with very similar characteristic frequencies. This is suspicious due to various \kepler\ instrumental effects (e.g. \citealt{edelson2014}, \citealt{kasliwal2015b}) and therefore we were motivated to analyze the case deeper. We tested the origin of these signals by performing the same PDS analysis using data from totally different objects in nature. Nearly all of our randomly selected planet candidates showed the same signals. Therefore, the instrumental origin of the signals is unambiguous.

\subsection{Comparison with Other Observations}

\subsubsection{The Case of ZW229-15}

Do these results imply that the discussions in \citet{dobrotka2017} focus on an instrumental artifact? As mentioned in Section~\ref{introduction}, one of the eight studied AGNs was observed by \swift. \citet{dobrotka2017} compared the \kepler\ PDS of this object with PDS calculated from \swift\ data. The X-ray light curve yielded very similar PDS with characteristic break frequencies in agreement with \kepler\ findings. This \swift\ detection manifests the reality of the signals, and the similarity with the \kepler\ data must be a very rare coincidence. However, our re-analysis of these \swift\ data using slightly different methods, which are able to suppress noise of the periodograms, and employed different binning, shows that at least the lower PDS component ($f_{\rm b1}$) has significantly lower value then derived previously, therefore does not match with the \kepler\ artifact. Thus the \swift\ detection likely represents a different signal at a close frequency, and the coincidence is not that glaring any more. The probability that the signal is real is at least 94\% based on red noise simulations. The higher frequency component in \swift\ data has non-negligible probability to be a result of simple red noise and poor light curve coverage. The similarity with the \kepler\ instrumental effect is also coincidental.

\begin{figure}
\resizebox{\hsize}{!}{\includegraphics[angle=-90]{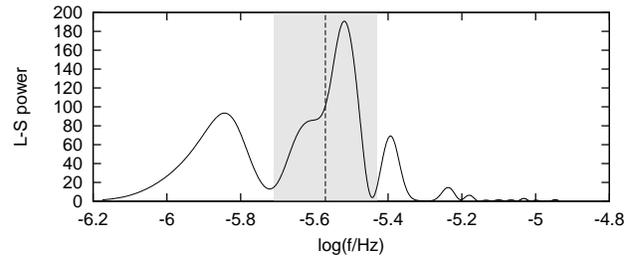}}
\caption{Lomb-Scargle periodogram of detrended \kepler\ light curve between days 717 and 734 (see bottom panel of Fig.~\ref{simul_lc_swift}). The vertical dashed line is the frequency of log($f$/Hz) = -5.57 derived from \swift\ PDS (see Table~\ref{pds_param}). The shaded arrea represents the error interval $\pm 0.14$.}
\label{l-s_zw229-15_kepler}
\end{figure}

The reality of the lower frequency component in the \swift\ PDS is supported by visible humps in \kepler\ data which are correlated with "flares" in the \swift\ light curve (already mentioned in \citealt{dobrotka2017}). In order to test the time scale of this (quasiperiodic) variability in \kepler\ data, we selected part of the light curve and fitted with a sinusoid plus linear function (blue solid line in the inset panel in the bottom of Fig.~\ref{simul_lc_swift}). The resulting fitted frequency is of log($f$/Hz) = -5.57, i.e. in exact match with the frequency derived from \swift\ PDS (Table~\ref{pds_param}). Moreover, the Fig.~\ref{l-s_zw229-15_kepler} represents the Lomb-Scargle periodogram of the light curve subsample between days 717 and 734 after detrending with 3$^{\rm th}$ order polynomial. The vertical line with the shaded arrea represents the frequency of log($f_{\rm b1}$/Hz) = $-5.57 \pm 0.14$ with error interval. The match is clear\footnote{Using the whole \kepler\ light curve with stronger detrending (20$^{\rm th}$ order polynomial) did not yield anything instructive}. This supports the reality of the $f_{\rm b1}$ signal in the \swift\ PDS, but also the reprocessing scenarion proposed by \citet{dobrotka2017}. Unfortunetely, the optical PDS equivalent of this signal in the \kepler\ data is buried in the instrumental pattern.

\subsubsection{Other Instruments and Objects}

There are also a variety of observations showing very similar values in the literature. \citet{mohan2014} studied fast variability of AGNs observed by \xmm. They fitted the PDSs with a broken power law and two observations of MRK\,335 and NGC\,3516 yielded break frequencies with values of log($f$/Hz) = $-4.22 \pm 0.41$ and $-4.24 \pm 0.25$. Furthermore, \citet{vaughan2003} studied PDS of MCG-6-30-15 observed by \xmm. They concluded that the PDS is well represented by a steep power law at high frequencies breaking to a flatter slope below log($f$/Hz) = -4.22. The authors used different models and one of them (see their Fig.~6 or Table~3) yield a break frequency of log($f$/Hz) = $-4.3^{+0.2}_{-0.1}$. Another AGN observed in X-rays was studied by \citet{mchardy2004}. The authors combined \xmm\ and \rxte\ data and in their Fig.~3 a break is noticeable around log($f$/Hz) = -5.1. However, this is just a visual estimate and must be taken with caution, because simple red noise can generate random features resembling the discussed PDS components. A similar visual inspection the PDSs of Ark\,564 and Ton\,S180 in Fig.~6 of\citet{edelson2002} shows a significant trend change at frequencies with approximate values of log($f$/Hz) = -4.82 ($\sim 1.5 \times 10^{-5}$\,Hz) and -4.7 ($\sim 2 \times 10^{-5}$\,Hz), -5.15 ($\sim 7 \times 10^{-6}$\,Hz) in Ark\,564 and Ton\,S180 PDSs, respectively. However, the low resolution of the PDSs must be taken into account and suggests non-negligible uncertainty.

All of these values are similar to the detected \kepler\ PDS components, implying that such signals do exists in AGNs or are at least probable in the studied frequency range. This explains the coincidence between the \kepler\ and \swift\ detections in the case of ZW229-15, i.e. the discussed \kepler\ features are most likely artifacts, while the \swift\ detection has a real physical origin. However, observing these (real) PDS components is not easy for two reasons. First, the corresponding time scales ($\sim 50$ and $\sim 150$\,ks) are too long for observations with instruments like \xmm\ and too short for good coverage with instruments like \swift. Second, ground observations are not helpful due to day-night cycles. An example of this observational limitation is shown in Fig.~13 of \citet{lobban2018} showing PDS from \rxte\, \swift\ and \xmm. The PDS components discussed here lay in the observational gap. However, the authors fitted the PDS with a broken power law and the resulting breaks were at log($f$/Hz) = $-5.7^{+0.4}_{-1.1}$ (hard band), $-5.6^{+0.4}_{-0.6}$ (soft band) and $-5.4^{+0.8}_{-1.1}$ (whole energy range). The similarity with signals from other mentioned observations is clear again.

\subsection{The blazar W2R\,1926+42 case}

Another good example motivating this work is study of the blazar W2R\,1926+42 by \citet{sasada2017}. The authors analysed the averaged shot profile of the fast variability in the \kepler\ light curve and found a break frequency in the PDS around log($f$/Hz) = $-4.39^{+0.05}_{-0.06}$ associated with the central dominant spike of the averaged shot.

\begin{figure}
\resizebox{\hsize}{!}{\includegraphics[angle=-90]{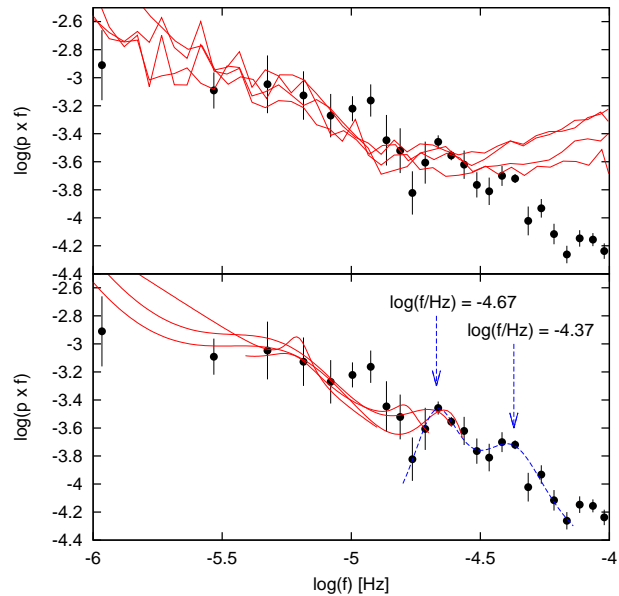}}
\caption{Upper panel - PDS (black points) of the blazar W2R\,1926+42 with four vertically adjusted AGN PDSs from Fig.~\ref{pds_all} (red solid lines) for direct comparison. Lower panel - The same as upper panel, but instead of AGN PDSs the corresponding fits are shown. The blue dashed line is the two component Lorentzian fit to the blazar PDS. The arrows show the resulting characteristic frequencies.}
\label{pds_blazar}
\end{figure}

To test whether this PDS pattern is of instrumental origin, we re-calculated the PDS for the same data using the method presented in Section~\ref{method}. The resulting PDS is shown in Fig.~\ref{pds_blazar}. The corresponding fits for all four AGNs from Fig.~\ref{pds_all} have been added and offset vertically for direct comparison (lower panel). Clearly, the behaviour of the PDS is different (upper panel), i.e. the white noise starts much earlier in the case of the four studied AGNs and a power excess (lower panel) at log($f$/Hz) = $-4.37 \pm 0.03$ ($\chi^2_{\rm red} = 1.04$, dof = 5) is clear, which does not coincide with the suspicious signal seen at log($f$/Hz) = $-4.67 \pm 0.06$. It is worth to noting that the PDS component at log($f$/Hz) = -4.37 is closer to the high frequency component of ZW229-15 found in the \swift\ light curve rather than in the \kepler\ data. Another power excess in the W2R\,1926+42 PDS is visible near log($f$/Hz) = -4.9.

Furthermore, \citet{sasada2017} studied obvious flare like events that are not seen in the eight \kepler\ AGNs, and the authors found a very similar shot profile to that found by \citet{negoro1994} in \ginga\ data of Cyg\,X-1, or Dobrotka et al. (in preparation) for \kepler\ data of MV\,Lyr but on much shorter time scales (in Cyg\,X-1 even shorter). These results indicate that the study by \citet{sasada2017} examined true astrophysical phenomena. If the PDS component at log($f$/Hz) = $-4.39^{+0.05}_{-0.06}$ is associated with the central spike of the averaged shot profile, the feature at log($f$/Hz) = -4.9 can correspond to longer substructures like the side-lobes.

\subsection{Conclusion}

While the \kepler\ data examined by \citet{dobrotka2017} are clearly contaminated by instrumental effects, the discussion and physics can be still applied to the revised \swift\ detection in ZW229-15. Moreover, it appears that some other AGN observations yield very similar PDS components, and the physical explanation in \citet{dobrotka2017} can be valid in these cases as well. However, as AGN show variability on time scales similar to the instrumental variabilities generated by \kepler, the data and interpretation of results require special techniques to effectively differentiate the signals.

\section{Summary}

The results of this work can be summarized as follows:

(i) We analyzed the four \kepler\ AGN light curves from \cite{revalski2014} and found the same PDS component frequencies as \citet{dobrotka2017}.

(ii) To investigate these components, we performed the same PDS analysis on five randomly selected \kepler\ field stars and found identical PDS features.

(iii) These results indicate that all of the detected PDS features are the result of uncorrected instrumental artifacts inherent to the \kepler\ instrument.

(iv) Based on observations with different telescopes like \xmm\, the X-ray light curves of AGNs do show real variability with comparable characteristic frequencies.

(v) The use of \kepler\ for variability and time-domain studies must be approached with caution to properly differentiate between instrumental artifacts and true astrophysical signals.

\section*{Acknowledgment}

AD was supported by the Slovak grant VEGA 1/0335/16 and by the ERDF - Research and Development Operational Programme under the project "University Scientific Park Campus MTF STU - CAMBO" ITMS: 26220220179.

M.R. gratefully acknowledges support from the National Science Foundation through the Graduate Research Fellowship Program (DGE-1550139).

We acknowledge the anonymous reviewer for helpful comments, particularly concerning the red noise simulations, which considerably improved this paper.

\bibliographystyle{mn2e}
\bibliography{mybib}

\label{lastpage}

\end{document}